\documentclass[twocolumn,english,aps,showpacs,superscriptaddress,amsmath,amssymb,tightenlines]{revtex4}
\usepackage{amsmath}
\usepackage{graphicx}
\usepackage{amssymb}

\makeatletter
\@ifundefined{textcolor}{}
{%
 \definecolor{BLACK}{gray}{0}
 \definecolor{WHITE}{gray}{1}
 \definecolor{RED}{rgb}{1,0,0}
 \definecolor{GREEN}{rgb}{0,1,0}
 \definecolor{BLUE}{rgb}{0,0,1}
 \definecolor{CYAN}{cmyk}{1,0,0,0}
 \definecolor{MAGENTA}{cmyk}{0,1,0,0}
 \definecolor{YELLOW}{cmyk}{0,0,1,0}
 }

\makeatother

\usepackage{babel}

\begin{document}

\title{Doubly excited ferromagnetic spin-chain as a pair of coupled kicked
rotors}

\author{T. Boness}

\affiliation{Department of Physics and Astronomy, University College London, Gower
Street, London WC1E 6BT, United Kingdom}

\author{K. Kudo}

\affiliation{Division of Advanced Sciences, Ochadai Academic Production, Ochanomizu
University, 2-1-1 Ohtsuka, Bunkyo-ku, Tokyo 112-8610, Japan}

\author{T.S. Monteiro}

\affiliation{Department of Physics and Astronomy, University College London, Gower
Street, London WC1E 6BT, United Kingdom}

\date{\today}
\begin{abstract}
We show that the dynamics of a doubly-excited Heisenberg spin-chain,
subject to short pulses from a parabolic magnetic field may be analyzed
as a pair of quantum kicked rotors. By focusing on the two-magnon
dynamics in the kicked XXZ model we investigate how the anisotropy
parameter - which controls the strength of the magnon-magnon interaction
- changes the nature of the coupling between the two ``image''
coupled Kicked Rotors. We investigate quantum state transfer possibilities
and show that one may control whether the spin excitations are transmitted
together, or separate from each other. 
\end{abstract}


\maketitle

\section{Introduction}

Over recent years, there has been sustained interest in coupled quantum
systems. Numerous studies investigated the causes and effects of decoherence
on a subsystem as it becomes entangled with its environment; others
probed the generation of bipartite entanglement between a pair of
quantum systems. It is vitally important to understand these processes
so they can be accounted for in protocols for quantum computation
and communication.

Studies of decoherence also shed light on the emergence of classical
behavior from quantum dynamics \cite{Zurek03,Benenti07}. Quantum
systems with a chaotic classical limit often feature in such studies.
For example, they can play the role of the environment: a 1D system
which displays chaos can replace a many-body heat bath (often modeled
by an infinite collection of quantum Harmonic Oscillators) as a source
of decoherence \citep{rossini:036209}. Other studies focused on entanglement
generation: the rate of growth of the von Neumann entropy of the subsystem
- ie the rate at which the subsystem becomes entangled with its environment
- is directly related to measures of the chaos in the subsystem's
classical limit \cite{Zurek94,Miller1999,Miller1999a}.

The chaos paradigm known as the Quantum Kicked Rotor (QKR) \cite{casati1979}
plays a central role in these studies. The QKR corresponds to the
dynamics of independent quantum particles evolving under the rather
simple Hamiltonian $H_{i}=\frac{p_{i}^{2}}{2}+K\sin x_{i}\sum_{n}\delta(t-nT)$,
where $K$ represents the kick-strength and $T$ the kick-period.
Cold atoms in pulsed standing waves of light were found to provide
a very clean realization of the QKR: in 1995, the phenomenon of the
quantum suppression of classical chaotic diffusion was clearly demonstrated
experimentally \cite{Moore}; later, the recovery of the classical
diffusive behavior in the presence of decoherence was also observed
\cite{Deco}. These works were followed by other studies by different
experimental cold-atom groups worldwide \cite{kickatom} probing wide-ranging
aspects of the QKR dynamics. In a previous work \cite{Boness061,Boness062},
we proposed that the singly-excited Heisenberg spin-chain in a pulsed
parabolic field could provide an exact physical realization of the
QKR: the dynamics of the spin-waves are given by a time-evolution
operator of analogous form to that of the QKR.

{\em Coupled} QKRs, have also been investigated in a number of
theoretical studies, though, unfortunately, no physical realization
has yet been achieved. In this case, one considers two QKR Hamiltonians
with an additional coupling potential $V$, i.e. $H=H_{1}+H_{2}+V(x_{1},x_{2},t)$.
In \cite{Fujisaki2003}, interactions which depended on the separation
of two rotors with a non-local sinusoidal term were investigated;
in \cite{Park2003} the two particles were confined to within a short
distance of each other. However, several studies considered a sinusoidal
coupling term dependent on a center-of-mass coordinate, \cite{LAK2001,Wood1990,Nag}
such as e.g.: $V(x_{1},x_{2},t)=K_{12}\cos\left(x_{1}+x_{2}\right)\sum_{n}\delta\left(t-nT\right)$.

In this work we show for the first time that the doubly-excited Heisenberg
spin-chain system may--to a good approximation--be analyzed as pair
of coupled kicked rotors. In fact, in this system, Nature even provides
a coupling term of the centre-of-mass form $K_{12}\cos\left(x_{1}+x_{2}\right)$.
The mapping is, however, far less straightforward than that found
for the one-excitation system in \cite{Boness061,Boness062}: the
coupling here is mediated by bound-pair eigenstates (not found in
the corresponding one-excitation chain), rather than spin waves, so
acts only over a restricted part of the ``image''
phase-space. The wavenumbers of the bound states are complex, further
complicating the mapping. Nevertheless, the analogy holds sufficiently
well, so one can use it to shed insight on the dynamics. Further,
it points to useful applications in state transfer, since we can use
this understanding to control whether the two spin-flips propagate
along the chain together, or separately. This adds to other applications
that make use of the single-excitation correspondence \citep{Gong2007}.

In Section II we review briefly the one-particle dynamics of the Heisenberg
XXZ spin chain and its mapping to the QKR. In section III we consider
the doubly-excited spin chain. We summarize essential features of
the well-known XXZ eigenstates and their dependence on the anisotropy
parameter $\Delta$. We then introduce the analogy with the two-particle
coupled QKR and explore the dynamics when the initial state consists
of two neighboring spin-flips. We also highlight two cases where the
kicked rotor correspondence simplifies: i) when $\Delta=0$ the kicked
spin-chain can be mapped to a pair of independent QKRs; and ii) when
$\Delta\gg1$ the bound states effectively trap two excitations on
neighboring sites and we show that in this limit, these bound states
give rise to a further analogy with the QKR. We finish, in section
IV with examples of how we can use these results to manipulate correlations
in the spin-flip locations.

\section{The Heisenberg spin-chain and its one-particle image}

The well-known spin-$1/2$ Heisenberg XXZ chain is governed by the
Hamiltonian:\begin{small} \begin{equation}
H_{hc}=-\frac{J}{4}\sum_{n=1}^{N}\left(2(\sigma_{n}^{+}\sigma_{n+1}^{-}+\sigma_{n}^{-}\sigma_{n+1}^{+})+\Delta\sigma_{n}^{Z}\sigma_{n+1}^{Z}\right)-B\sum_{n}\sigma_{n}^{Z}.\label{eq:Hhc}\end{equation}
\end{small}When investigating the dynamics of spin-chains such as
this, a useful approach is to invoke quasi-particle models and interpret
excited states as systems of indistinguishable particles. In some
cases, it is even possible to map the dynamics to a one-body ``image''
system which approaches the classical limit as $N\to\infty$ \cite{Prosen}.

$H_{\textrm{hc}}$ conserves the number of spin-flips and a single
excitation represents a spin-wave which distributes a single spin-flip
throughout the chain. Higher excited-states correspond to multiple
spin-waves which interact when they coincide through both an exclusion
process (no two spin flips can simultaneously occupy the same site)
as well as a mutual interaction induced by the $\sigma_{n}^{z}\sigma_{n+1}^{z}$
(Ising) term - the strength of which is determined by the anisotropy
parameter $\Delta$. Note that $\Delta=0$ corresponds to the XX0
chain and $\Delta=1$ is the isotropic Heisenberg chain.

The eigenstates for the single spin-flip sector of \eqref{eq:Hhc},
spanned by the basis states $\{|n\rangle=\sigma_{n}^{-}|\uparrow\uparrow\ldots\uparrow\rangle:n=1,\ldots,N\}$,
are translationally invariant magnon states with momenta $\kappa$:
\begin{equation}
|\kappa\rangle=\frac{1}{\sqrt{N}}\sum_{n=1}^{N}e^{i\kappa n}|n\rangle.\label{eq:magnon}\end{equation}
Note that periodic boundary conditions are used, i.e. the configuration
is a closed ring with $\sigma_{j+N}^{\alpha}=\sigma_{j}^{\alpha}$.
The magnon momenta are determined by these conditions and take the
values $\kappa=2\pi I/N$, $I=1,...,N$. These states have energy:
\begin{equation}
E-E_{0}=2B+J(\Delta-\cos\kappa),\label{dispersion}\end{equation}
where the ground state energy $E_{0}=-J\Delta N/4$, i.e. $H_{hc}|\uparrow\uparrow\ldots\uparrow\rangle=E_{0}|\uparrow\uparrow\ldots\uparrow\rangle$.
Adding an external parabolic kicking field to the Heisenberg Hamiltonian
gives: \begin{equation}
H=H_{hc}-\frac{B_{Q}}{4}\sum_{n=1}^{N}(n-n_{0})^{2}\sigma_{n}^{Z}\sum_{j=1}\delta(t-jT),\label{eq:HKHSC}\end{equation}
where the kicking field has amplitude $B_{Q}$ with minimum at $n_{0}$.
Time evolving the time-periodic $H$ for one period $T$ yields a
unitary map, \begin{equation}
|\psi(t=(j+1)T)\rangle=U(T)|\psi(t=jT)\rangle\end{equation}
 where \begin{equation}
U(T)=e^{-iTH/\hbar}=e^{i\frac{B_{Q}}{4\hbar}\sum_{n=1}^{N}\left(n-n_{0}\right)^{2}\sigma_{n}^{Z}}e^{-iTH_{hc}/\hbar}\label{eq:FLKXXZ}\end{equation}
since the $\delta$-kick nature of the time-dependent field permits
us to split the operators. 

Using \eqref{eq:magnon} and \eqref{dispersion}, it was shown in
\cite{Boness061,Boness062}, that the matrix elements of $U(T)$ in
the single-flip basis $\{|n\rangle\}$ have a form very similar to
the matrix used to evolve the quantum chaos paradigm, the QKR. These
are given by: \begin{equation}
U_{nn'}\simeq e^{i\frac{B_{Q}}{2}(n-n_{0})^{2}}i^{n'-n}J_{n'-n}(JT)\label{eq:spinprop}\end{equation}
for the spin-ring (an analogous form was given in \cite{Boness061}
for open boundary conditions). Here, $J_{n}(x)$ denotes a Bessel
function of order n and, for convenience, we have set $\hbar=1$.

We recall the form of the QKR Hamiltonian: \begin{equation}
H_{QKR}=\frac{\hat{p}^{2}}{2}-k\cos\hat{x}\sum_{j=1}\delta(t-jT).\label{eq:HQKR}\end{equation}
with $x\in[0,2\pi)$. In its classical limit, the dynamics is described
by the famous Standard Map which is known to display a transition
from integrability to chaos as the \textit{Stochasticity Parameter},
$K=kT$, is increased. For $K\lesssim1$ diffusion in momentum is
blocked by invariant tori running through classical phase space. At
large $K$ phase space is almost completely chaotic and unbounded
diffusion in momentum is typically seen. However, in certain ranges
of $K\approx2j\pi,\, j\in\mathbb{Z},$ small transporting islands
known as ``Accelerator Modes'' (AM) appear in classical
phase space and give rise to anomalous diffusion. In the QKR, diffusion
of momentum at large $K$ is suppressed by quantum interference in
a process known as Dynamical Localization \cite{Fishman1982,Grempel1984}.
We can express the QKR time propagator in a basis of plane waves $|l\rangle=\exp(ilx)$:
\begin{equation}
\langle l|U_{QKR}(T)|l'\rangle=e^{-il^{2}\tau/2}i^{l'-l}J_{l'-l}\left(\frac{K}{\tau}\right).\label{eq:propqkr}\end{equation}
Here $\tau=\hbar T$ is the rescaled effective Planck's constant.

Comparing the above with \eqref{eq:spinprop} we see that the QKR
and spin-chain propagators are of similar form, provided we identify
$\frac{K}{\tau}\to JT$ and note that the kicking field $B_{Q}\to\tau$
plays the role of an effective Planck's constant. In effect, the spin-chain
equivalent to the $K\cos x$ term in $H_{\textrm{QKR}}$ arises from
the dispersion relation of the spin-waves, ie re-writing \eqref{eq:Hhc}:
\begin{equation}
H_{hc}=\sum_{\kappa}J(\Delta-\cos\kappa)|\kappa\rangle\langle\kappa|\label{eq:disp1}\end{equation}
So, to make the QKR $\to$ spin-wave mapping we also had to identify
position ($x$) in the QKR, with momentum in the spin-chain ($\kappa$);
and momentum in the QKR with position (spin-site) in the chain.

With the aid of this mapping we can identify the spin-wave equivalent
of classical chaos phenomena such as Accelerator Modes (AM), transport
on tori \cite{Kudo}, cantori or stable islands, and quantum chaos
phenomena like Dynamical Localization \cite{casati1979}. The classical
transporting islands represented by the AM have evident potential
applicability in quantum state transfer so, below, we investigate
these in particular: they occur for $K\approx2j\pi$ where $j$ is
an integer. In the classical image phase-space, they correspond eg
to initial conditions located around $(x_{0},p_{0})\approx(\pm\pi/2,0)$
which {}``hop'' in momentum each period such that at $t=nT$ ($n\in\mathbb{Z}$):\begin{equation}
(x_{n},p_{n})\approx(\pi/2,\mp2\pi nj).\end{equation}
Quantum mechanically, if the effective Planck's constant is small
enough, these islands can support Gaussian states that follow the
classical trajectories - i.e. they ``hop'' in momentum every period
\citep{Hanson}. Gaussian excitations were indeed seen in the one-flip
spin-chain, \citep{Boness061}, provided the initial spin-flip occurs
at a site near the minimum of the magnetic kicking field and $K=JTB_{Q}\approx2\pi j$.
The excitations were seen to hop approximately $2\pi/B_{Q}$ sites
each period, with little dispersion.

We now consider the two-flip case.

\section{Two spin excitations}

\subsection{Bound-pair states and spin-waves}

Eigenstates in the double excitation sector are expressed, via the
Bethe ansatz, as pairs of spin waves \citep{KarbMull98}: \begin{equation}
|\kappa_{1},\kappa_{2}\rangle=A(\kappa_{1},\kappa_{2})\sum_{0\leq n_{1}<n_{2}\leq N}a(n_{1},n_{2})\,|n_{1},n_{2}\rangle.\label{eq:ges}\end{equation}
 where $A(\kappa_{1},\kappa_{2})$ is a normalization constant. $|n_{1},n_{2}\rangle$
denotes a state with a spin-flip at sites $n_{1}$ and $n_{2}$. Bethe's
ansatz for the amplitude is \begin{equation}
a(n_{1},n_{2})=e^{i(\kappa_{1}n_{1}+\kappa_{2}n_{2}+\theta/2)}+e^{i(\kappa_{1}n_{2}+\kappa_{2}n_{1}-\theta/2)}.\label{eq:bamp}\end{equation}
 The scattering phase $\theta(\kappa_{1},\kappa_{2})$ accounts for
the interaction between the pair of spin-waves. On applying $H_{hc}$
in \eqref{eq:Hhc} to these states and solving the eigenvalue equations,
one obtains the dispersion relation \begin{equation}
E-E_{0}=4B+J\left(2\Delta-\cos\kappa_{1}-\cos\kappa_{2}\right)\label{eq:2mnrg}\end{equation}
 and also a relation between $\theta$ and the quasi-momenta, the
Bethe Ansatz Equation (BAE): \begin{equation}
e^{i\theta}=-\frac{1+e^{i\left(\kappa_{1}+\kappa_{2}\right)}-2\Delta e^{i\kappa_{1}}}{1+e^{i\left(\kappa_{1}+\kappa_{2}\right)}-2\Delta e^{i\kappa_{2}}}\label{eq:bae}\end{equation}
 Further restrictions are imposed by the periodic boundary conditions:
\begin{equation}
N\kappa_{1}=2\pi\lambda_{1}+\theta,\,\,\,\, N\kappa_{2}=2\pi\lambda_{2}-\theta.\label{eq:keq}\end{equation}
where the Bethe quantum numbers $\lambda_{1}\leq\lambda_{2}$ are
integers in the range $\lambda_{i}\in\{0,1,\ldots,N-1\}$. By solving
the coupled system of equations in \eqref{eq:bae} and \eqref{eq:keq},
$\kappa_{1,2}$ and $\theta$ can be obtained. Broadly speaking, these
solutions fall into two groups depending on whether $\theta$ has
an imaginary component. The majority of the solutions of \eqref{eq:bae}
are real - these correspond states of two magnons which scatter off
each other. For $\Delta=0$ all the available solutions of \eqref{eq:bae}
are real (and equal to $\pi$). When $\theta$ is complex, the eigenstates
correspond to bound states of two spin flips. The probability amplitudes
of these states are at a maximum when the flips are on neighboring
sites and they decay exponentially with the separation of the flips.
While for any given $\Delta\neq0$ the widths of these states vary
with the total momentum $\kappa_{1}+\kappa_{2}$, they become narrower
as $\Delta$ increases. Crucially, for long chains ($N\to\infty$)
the energy of these states can be written \cite{Takahashi}: \begin{equation}
E-E_{0}=4B+J\Delta-\frac{J}{2\Delta}(1+\cos(\kappa_{1}+\kappa_{2})).\label{eq:string_disp}\end{equation}
 when $\Delta>0$. %
\begin{figure*}
\includegraphics[scale=0.45]{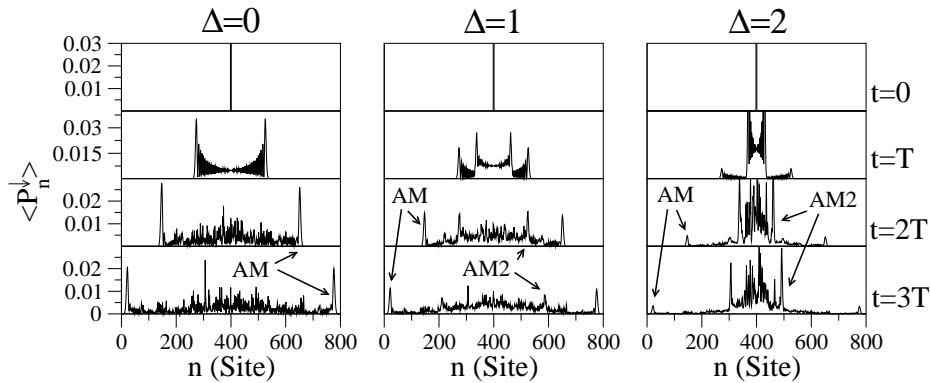} \caption{Showing the production of {}``bound state'' accelerator modes (AM2)
which move slowly, and fast ``scattering state'' accelerator modes
(AM). When $\Delta=0$ only the AM are present, in this case the dynamics
maps to two independent QKRs. With increasing $\Delta$ the AM2 become
dominant. We plot the on-site magnetization for for two initially
neighboring spin-flips $|\psi(0)\rangle=|400,401\rangle$. We have
chosen here, the parameters $K_{s}=13$, $B_{Q}=0.1$, $n_{0}=400$
and $T=1$ for chains of $800$ spins, which are known to produce
Gaussian excitations in a singly-excited chain. }

\label{fig:fig1} 
\end{figure*}

\subsection{Analogy with a pair of coupled kicked rotors}

The departure point for our analysis of the spin dynamics as a system
of coupled QKRs is the two-excitation spin-Hamiltonian, equivalent
of (\ref{eq:disp1}): \begin{small} \begin{equation}
\begin{array}{ccl}
H_{hc} & = & {\hat{P}_{s}}\sum_{\kappa_{1,2}}J\left(\Delta-\cos(\kappa_{1})-\cos(\kappa_{2})\right)|\kappa_{1},\kappa_{2}\rangle\langle\kappa_{1},\kappa_{2}|\\
 &  & -{\hat{P}_{b}}\sum_{\kappa_{1,2}}\frac{J}{2\Delta}\left(1+\cos(\kappa_{1}+\kappa_{2})\right)|\kappa_{1},\kappa_{2}\rangle\langle\kappa_{1},\kappa_{2}|\\
 &  & +\mathbb{\hat{I}}\left(E_{0}+4B+\Delta J\right)\end{array}\label{eq:HHCdiag}\end{equation}
 \end{small} for $\Delta>0$. Here ${\hat{P}_{s}}$ is a projector
on to the scattering-state component of Hilbert space, and ${\hat{P}_{b}}$
onto the bound states.

Comparing the above with the typical coupled QKR potential $V(x_{1},x_{2})=K_{1}\cos x_{1}+K_{2}\cos x_{2}+K_{12}\cos(x_{1}+x_{2})$
and identifying $\kappa_{i}\to x_{i}$ and $K_{12}\to\frac{JT}{2\Delta}$
might suggest that the scattering states be interpreted as giving
rise to a pair of kicked rotors; and that a coupling between these
rotors arises due to the bound states. However, we note the important
difference that the $\kappa_{i}$ for the scattering and bound states
correspond to complementary portions of the ``image''
phase space. For the bound states, $\kappa_{i}$ is complex, but $(\kappa_{1}+\kappa_{2})$
is real. In addition, we show below (in \eqref{eq:NNQKRME}) that
in fact, for large $\Delta$, $K_{12}$, i.e. the effective coupling
is twice as large as suggested by \eqref{eq:HHCdiag}.

The parabolic kick will couple the eigenstates to each other (including
coupling bound-pair and scattering states). As $\Delta$ increases,
the overlap between the bound and scattering state energies decreases
and the two bands separate for $\Delta>2$. This will suppress the
coupling and imply that for large $\Delta$, if the initial state
has negligible overlap with the bound subspace, the dynamics will
be essentially uncoupled.

\subsection{Evolution of $|n,n+1\rangle$ initial states}

In this section, we explore the dynamics of an initial state prepared
with two spin-flips localized on neighboring sites near the center
of the chain $|\psi(0)\rangle\equiv|n_{0},n_{0}+1\rangle$. Parameters
corresponding to accelerator modes are used: $JT=130$ and $B_{Q}=1/10$,
so $K\sim4\pi$. Fig. \ref{fig:fig1} shows the resulting on-site
magnetization $\langle P_{n}^{\downarrow}\rangle=\langle\frac{1}{2}(1-\sigma_{n}^{Z})\rangle$
of $|\psi(0)\rangle$ after successive applications of the map \eqref{eq:FLKXXZ}
for $\Delta=0$, $1$ and $2$.

\begin{figure*}
\includegraphics[bb=30bp 0bp 843bp 294bp,clip,scale=0.62]{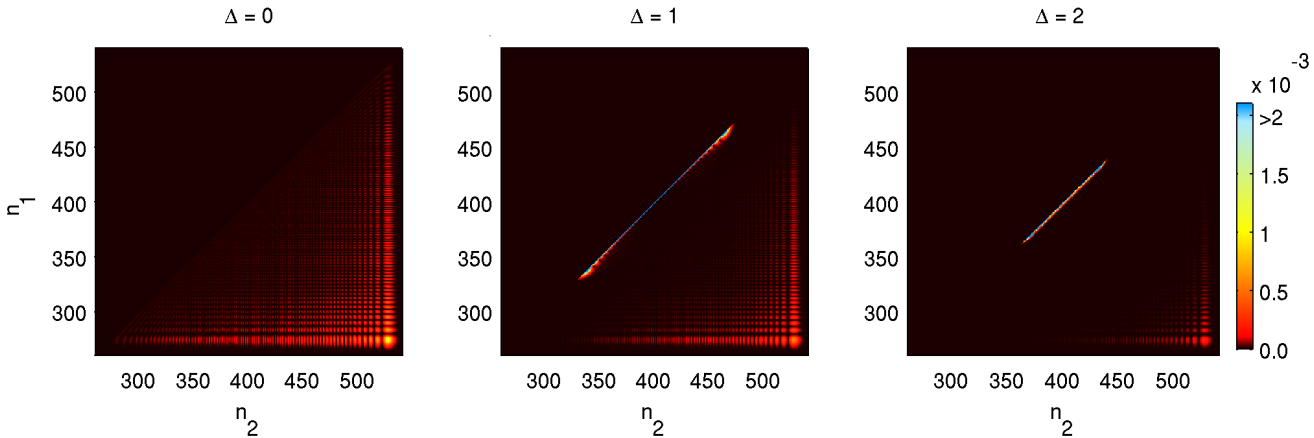}
\caption{(color online) Spin-spin correlations corresponding to Fig. \ref{fig:fig1}
at $t=T$. The two-site correlation function $\langle P_{n_{1}}^{\downarrow}P_{n_{2}}^{\downarrow}\rangle$,
equal to the probability of finding one flip on site $n_{1}$ and
the other on $n_{2}$, is shown. At $\Delta=0$ the spins are anti-correlated
in contrast to Fig. \ref{fig:fig1} which suggests the dynamics of
uncoupled particles. $\Delta=1$ and $2$ have an anti-correlated
component (flips separate) as well as an additional component where
the flips travel together.}

\label{fig:fig2} 
\end{figure*}

When $\Delta=0$, a pair of hopping wavepackets is produced. Each
travels $2\pi/B_{Q}\approx130$ sites each period. This is consistent
with the single-particle accelerator modes (see previous section).
In contrast, when $\Delta=1$ or $2$, there are two sets of hopping
wavepackets. One pair (AM) evolve like those in the $\Delta=0$ chain,
while the other pair (AM2) hop approximately $\pi/(\Delta B_{Q})$
sites each period. For $\Delta=1$ the AM2 wavepackets decay rapidly
and by the third period ($t=3T$) they are almost indistinguishable
from the chaotic central portion.

To get a more complete picture of the dynamics we plot, in Fig. \ref{fig:fig2},
the two-site correlation function $\langle P_{n_{1}}^{\downarrow}P_{n_{2}}^{\downarrow}\rangle$
for $|\psi(T)\rangle$, allowing us to follow the relative positions
of the spin-flips. We find that the AM2 wavepackets contain flips
that travel together, this suggests they are supported by the bound
states. The AM wavepackets on the other hand appear in an anti-correlated
portion of the wavefunction.

We now consider these two different kinds of behaviour in more detail.

\subsection{$\Delta=0$ and `independent' QKRs}

For $\Delta=0$ only the exclusion interaction is present between
flips. The effects of this interaction are subtle and sensitive to
the initial conditions. For certain cases, where the flips are initially
separated by an odd number of sites, it has been shown to change the
character of entanglement when the two excitations collide \citep{Amico04}.
A separate study on the transfer of entangled states in a doubly excited
XX0 chain was carried out in \citep{Sub04}.

Here, we are interested what influence the exclusion interaction has
on the QKR-like behaviour of excitations in the kicked spin-chain.
The $\Delta=0$ model can be mapped to a system of spinless fermions
via the Jordan-Wigner transformation (see appendix for details). The
number of fermions matches the number of spin-flips. The exclusion
interaction is accounted for by the anti-commuting property of the
fermionic operators. Consequently, the fermions are ``free'' (non-interacting)
and can be evolved separately under the single-particle dynamics.
The result of this is that the kicked chain maps to a system of non-coupled
QKRs. However, in the spin representation, the exclusion interaction
is still relevant (the spin-flips do not evolve separately). To see
this we make use of the Floquet operator in the two spin-flip basis
(see appendix): \begin{small} \begin{eqnarray}
\langle n_{1},n_{2}|U^{\Delta=0}(T)|m_{1},m_{2}\rangle=\qquad\qquad\qquad\qquad\qquad\notag\\
\qquad e^{-i\frac{B_{Q}}{2}\left[(n_{1}-n_{0})^{2}+(n_{2}-n_{0})^{2}\right]}i^{n_{1}+n_{2}-m_{1}-m_{2}}\times\qquad\notag\\
\qquad\left[J_{n_{1}-m_{1}}(\beta)J_{n_{2}-m_{2}}(\beta)-J_{n_{1}-m_{2}}(\beta)J_{n_{2}-m_{1}}(\beta)\right],\label{eq:XXTMPROP}\end{eqnarray}
\end{small} where $\beta=JT$ and $N\to\infty$.

The effects of the exclusion interaction are not actually seen in
Fig. 1. For example, the on-site magnetization after one period (i.e.
for $|\psi(t=T)\rangle$) is $\langle P_{n}^{\downarrow}\rangle=\sum_{n_{1}<n}|U_{n_{1},n,n_{0},n_{0}+1}^{\Delta=0}|^{2}+\sum_{n_{2}<n}|U_{n,n_{2},n_{0},n_{0}+1}^{\Delta=0}|^{2}$,
which is $\langle P_{n}^{\downarrow}\rangle=J_{n-n_{0}}^{2}(\beta)+J_{n-n_{0}+1}^{2}(\beta)$.
This is the same as for two independent spin-flips initialized at
sites $n_{0}$ and $n_{0}+1$. Using the free-fermion correspondence,
it is straightforward to show that for all later times $\langle P_{n}^{\downarrow}\rangle$
is exactly equivalent to the sum of expectations for a pair of uncoupled
QKRs. 

The coupling induced by the exclusion interaction is, however, evident
in Fig. 2, which plots the two-site correlations after the first period:
$\langle P_{n_{1}}^{\downarrow}P_{n_{2}}^{\downarrow}\rangle=|U_{n_{1},n_{2},n_{0},n_{0}+1}^{\Delta=0}|^{2}$.
Its effect, for this particular initial state, is to prevent the spin-flips
from travelling together. The two site correlation is highest when
the flips travel $JT=130$ sites in opposite directions. If the flips
were non interacting (i.e. allowed to co-exist on the same site) then
it would be equally likely the flips would travel together or apart.
Different correlations are seen when the initial separation of the
flips is changed.

So when $\Delta=0$, where the Heisenberg chain eigenstates consist
entirely of scattering states, the behavior of two spin-flips is like
that of two Kicked Rotors except the flips build up correlations in
their relative positions.

\subsection{Bound State QKRs for large $\Delta$}

When $\Delta\neq0$, the spin-flips can form bound pairs, and when
$\Delta=1$ or $2$, a pair of neighbouring spin-flips will overlap
with both bound and scattering eigenstates. The additional features
seen in the probability distributions when $\Delta=1\textrm{ and }2$
are remnants of QKR-like behavior of bound states that appears in
the limit of large $\Delta$. In this limit, the bound states confine
the flips to neighboring sites. Santos and Dykman \citep{Santos03}
use a perturbation expansion in spin coupling strength $J$ to produce
an effective Hamiltonian when $\Delta\gg1$. In this approximation
the bound state amplitudes are: \begin{equation}
a(n_{1},n_{2})=\delta_{n_{1},n_{2}-1}e^{i(\kappa_{1}+\kappa_{2})n_{1}}\label{eq:bs_larged}\end{equation}
and their dispersion relation remains unchanged from \eqref{eq:string_disp}.
Clearly, in center of mass coordinates, the bound states have the
same form as a single magnon solution. Naturally, this similarity
extends to the dynamics of states on the nearest neighbor (NN) subspace,
$\{|n,n+1\rangle\}$: Two initially neighboring spin-flips evolve
together in approximately the same way as a lone flip in the single
excitation basis $\{|n\rangle\}$. We anticipate that for $\Delta\gg1$:
\begin{equation}
\langle n_{1},n_{2}|U_{hc}(t)|m,m+1\rangle\approx i^{n_{1}-m}J_{n_{1}-m}\left(\frac{Jt}{2\Delta}\right)\delta_{n_{1},n_{2}-1}\label{eq:nnmeres}\end{equation}
 where the propagation of the neighboring flips is slower than for
a single flip - it is scaled by $J/(2\Delta)$ rather $J$.

The influence of the kicking field on the NN subspace can be incorporated
into \eqref{eq:nnmeres} to give: \begin{equation}
\begin{array}{lll}
\langle n_{1},n_{2}|U_{H}(T)|m,m+1\rangle & \approx & \delta_{n_{1},n_{2}-1}e^{iB_{Q}(n_{1}-(n_{0}-\frac{1}{2}))^{2}}\\
 &  & \qquad\quad i^{n_{1}-m}J_{n_{1}-m}\left(\frac{JT}{2\Delta}\right)\end{array}\label{eq:NNQKRME}\end{equation}
 up to a global phase. Again, we see an analogy to a QKR, with stochasticity
parameter $K_{b}=JTB_{Q}/\Delta$ and effective Planck's constant
$\tau_{b}=2B_{Q}$.

\begin{figure}
\includegraphics[scale=0.7]{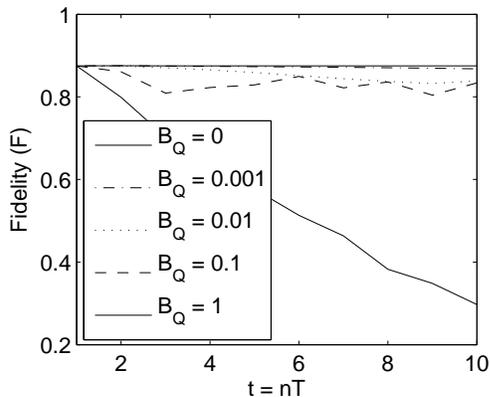} \caption{Showing the decay, over time, of the correspondence between the dynamics
of nearest neighbor flips and a QKR. $F$ is the fidelity of the time
evolution of two spin-flips intially on neighboring sites $|\psi(t=nT)\rangle=[U(T)]^{n}|100,101\rangle$
to the matrix elements \eqref{eq:nnmeres} and \eqref{eq:NNQKRME}
for various $B_{Q}$ and $JT/\Delta=5$, $\Delta=2$.}

\label{nnqkrfid} 
\end{figure}

We expect the accuracy of this approximation to fall with decreasing
$\Delta$ as the bound states become broader and are coupled more
strongly to the scattering states by the kicking field. However, we
show in Fig. \ref{nnqkrfid} that even for $\Delta=2$, QKR-like behavior
is still seen on the NN subspace for short times.

\subsection{Scattering State QKRs and Center of Mass Diffusion}

We now consider parameter ranges for which a single particle displays
Dynamical Localization. Taking $K=JTB_{Q}=5.0$ and $B_{Q}=1$ a lone
flip initially spreads diffusively but at long times this spreading
saturates and the flip becomes exponentially localized $\langle P_{n}^{\downarrow}\rangle\sim\exp\{-2|n-n_{\textrm{init}}|/L\}$
with a localization length $L=(JT)^{2}/4$. The diffusion time is
usually increased for coupled kicked rotors, e.g. in a related study
\cite{Shepelyansky94} a pair of rotors coupled locally in momentum
$U\delta{p_{1},p_{2}}$ were shown to localize with a much greater
$L$.%
\begin{figure}
\includegraphics[scale=0.6]{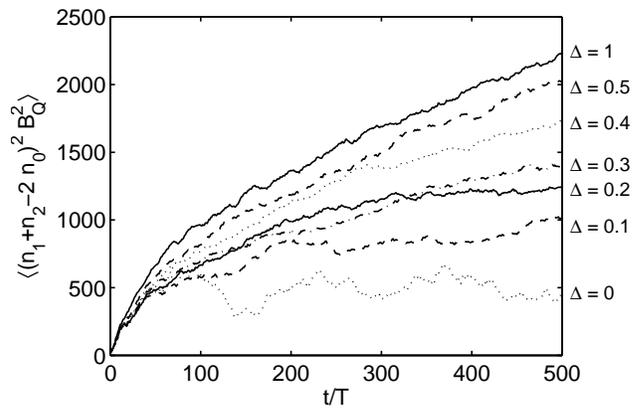} \includegraphics[scale=0.62]{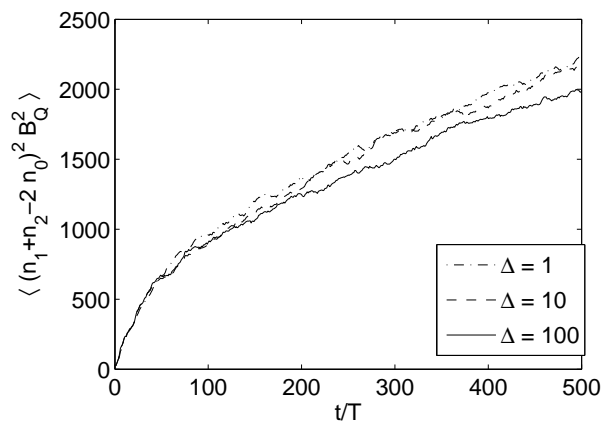}
\caption{Influence of the $\sigma_{i}^{z}\sigma_{i+1}^{z}$ coupling on the
growth of the `center of mass' second moments for two flips initialized
10 sites apart and parameters $K=5$ and $B_{Q}=1$ on a chain of
$200$ spins. }

\label{fig:diff} 
\end{figure}

In Fig. \ref{fig:diff} we follow the center of mass spreading of
two spin-flips using the second moment $\langle(n_{1}+n_{2}-2n_{0})^{2}B_{Q}^{2}\rangle$.
The flips are initialized 10 sites apart so in the limit of large
$\Delta$ this state should overlap only with scattering states. The
spin-distribution localizes for $\Delta=0$ as expected for an uncoupled
QKR; however, for larger $\Delta$, the diffusion is not halted, but
slows down appreciably after the ``break-time'' at $\Delta=0$.
This slower diffusion saturates and reaches a constant rate for $\Delta\gtrsim1$.

For large $\Delta$, due to the large energy gap, the kicking field
will not significantly couple the bound and scattering states so the
quantum state is supported only by the scattering states for all time.
The behavior of the diffusion however, does not reduce to that of
uncoupled kicked rotors (as might be suggested by the dispersion relation
in \eqref{eq:HHCdiag}). This is because the scattering states do
not exactly reduce to a pair of magnons-even in the large $N$ limit
where the corrections to the momenta $\kappa_{1/2}$ (see \eqref{eq:keq})
vanish (i.e. $\theta/N\to0$). They are distorted by the Ising term
and correspondingly the presence of the bound states. This can be
seen by rearranging the Bethe equations in section A. For $\Delta\neq0$
the scattering state amplitudes can be written\begin{small}\begin{eqnarray}
a(n_{1},n_{2}) & \propto & e^{i\frac{\kappa_{c}}{2}\left(n_{1}+n_{2}\right)}\left[\sin\left(\kappa_{r}\left(n_{1}-n_{2}+1\right)\right)\right.\nonumber \\
 &  & \left.-\frac{\cos\frac{\kappa_{c}}{2}}{\Delta}\sin\left(\kappa_{r}\left(n_{1}-n_{2}\right)\right)\right]\end{eqnarray}
\end{small}where $\kappa_{c}=2\pi\left(\lambda_{1}+\lambda_{2}\right)/N$
and $\kappa_{r}=\left(\pi\left(\lambda_{1}-\lambda_{2}\right)+\theta\right)/N$.
Notably, when $\Delta\gg1$, the scattering states typically have
no overlap with the $|n,n+1\rangle$ subspace (except, eg, for the
$\lambda_{1}=\lambda_{2}=\theta=0$ state) as this is occupied by
the bound states.

\section{Possibilities for controlling the evolution of spin-flips}

These results may be of interest in the context of quantum information
and state transfer as they suggest possibilities for manipulating
the evolution of spin-flips (and spin-correlations) in a Heisenberg
chan. Clearly, the evolution of a two-particle state on the non-kicked
XXZ chain depends on $\Delta$ and the shape of the initial wavepacket.
These two factors also determine the proportion of the wavefunction
that is supported by the scattering/bound states. In the kicked chain,
for large $\Delta$, the scattering and bound states correspond to
different QKR images: The bound ($b$) and scattering state ($s$)
QKR image parameters are related via $K_{b}=K_{s}/\Delta$ and $\tau_{b}=2\tau_{s}$.
So, by picking suitable values of $JT$ and $B_{Q}$ we can select
which dynamical regimes the bound and scattering components correspond
to.

For example, one could halt the propagation of either the bound or
scattering state portion of the wavefunction and allow the rest to
travel. A possible way to do this is to make use of resonances in
the QKR. These occur for $\tau=4\pi r$ where $r$ is rational. For
$r=1$ (primary resonance) ballistic spreading occurs in momentum
for the QKR (position for the spin chain) and when $r=1/2$ (antiresonance)
diffusion in momentum can be suppressed. These two conditions could
be achieved for the bound and scattering states respectively by setting
$\tau_{b}=4\pi$. This would lead to ballistic diffusion for initially
neighboring flips and could prevent flips that are initially well
separated from spreading.

\section{Conclusions}

We have investigated the dynamics of a pair of spin-flips on a periodically
kicked Heisenberg chain, focusing on the roles of the scattering and
bound eigenstates of the underlying time independent model. Analogies
to coupled and independent rotor systems have been identified and
analysed.

T. B. acknowledges support from the EPSRC. This work was partly supported
by Grant-in-Aid by MEXT, Japan.

\appendix

\section{Time Evolution for the Kicked XX0 Model}

Here, we show how the kicked XX0 chain (obtained from \eqref{eq:HKHSC}
by setting $\Delta=0$) maps to a system of independent QKRs. This
is done by first applying the Jordan-Wigner transformation \cite{Jordan},
a non-local mapping of spin-flips on the chain to free fermions on
a lattice. This transformation defines fermion creation and annihilation
operators, $c_{j}^{\dagger}$ and $c_{j}$ respectively, in terms
of spin operators so that $\sigma_{i}^{z}=(1-2c_{i}^{\dagger}c_{i})$
and {\small \begin{equation}
\hat{\sigma}_{i}^{+}=\left[\prod_{j<i}(1-2c_{j}^{\dagger}c_{j})\right]c_{i},\quad\hat{\sigma}_{i}^{-}=\left[\prod_{j<i}(1-2c_{j}^{\dagger}c_{j})\right]c_{i}^{\dagger}.\label{eq:JWtrans}\end{equation}
}The product of $(1-2c_{j}^{\dagger}c_{j})$ terms accounts for the
difference between inter-particle exchange statistics - negative for
fermions and positive for spin-flips. The creation and annihilation
operators obey the standard commutation relations: $\{c_{i},c_{j}^{\dagger}\}=\delta_{i,j}$
and $\{c_{i},c_{j}\}=0$ and are defined with respect to a vacuum
state $|\phi\rangle$ such that $c_{j}|\phi\rangle=0$. 

Making use of the transformation and setting $\Delta=0$, the kicked
spin-chain Hamiltonian \eqref{eq:HKHSC} becomes: \begin{small} \begin{equation}
\begin{array}{ccl}
H & = & -\frac{J}{2}\left[\sum_{j=1}^{N-1}\left(c_{j}^{\dagger}c_{j+1}+c_{j+1}^{\dagger}c_{j}\right)-(-1)^{r}\left(c_{1}^{\dagger}c_{N}+c_{N}^{\dagger}c_{1}\right)\right]\\
\\ &  & +\ \frac{B_{Q}}{2}\sum_{j=1}^{N}(j-j_{0})^{2}(1-2c_{j}^{\dagger}c_{j})\ \delta\left(t/T\right).\end{array}\label{eq:H_kf}\end{equation}
\end{small}The transformed Hamiltonian has boundary terms that depend
on whether the number of fermions, $r$, is odd or even; these arise
from the periodic boundary conditions $\sigma_{N+1}^{\pm}=\sigma_{1}^{\pm}$
and $\sigma_{N+1}^{Z}=\sigma_{1}^{Z}$. 

We now calculate the result of time evolving over one period. Using
the Heisenberg picture, we define $c_{j}^{\dagger}(T)=U^{\dagger}(T)c_{j}(0)U(T)$,
where $U(T)$ is the Floquet operator of eq. \ref{eq:FLKXXZ}. The
absence of any mutual interaction (i.e. terms of the form $v(j_{1}j_{2},j_{1}^{'},j_{2}^{'})c_{j_{1}}c_{j_{2}}c_{j_{1}^{'}}^{\dagger}c_{j_{2}^{'}}^{\dagger}$)
in \eqref{eq:H_kf} implies the fermions are {}``free'' and therefore
the fermion operators can be time evolved using single particle states:
$U^{\dagger}(T)c(\psi(0))U(T)=c^{\dagger}(\psi(-t))$ where $c^{\dagger}(\psi(-t))$
creates the single particle state $\psi(-t)=U^{\dagger}(T)\psi(0)$.
This corresponds to the single-flip basis, so $c_{j}^{\dagger}(T)|\phi\rangle=U^{\dagger}(T)|j\rangle$.
Using the matrix elements $U_{nn'}$ in \eqref{eq:spinprop} gives,
up to a global phase,\begin{equation}
c_{j}^{\dagger}(T)\approx e^{iB_{Q}(j-j_{0})^{2}/2}\sum_{j'}i^{j'-j}J_{j'-j}(-JT)c_{j'}^{\dagger}(0).\label{eq:fevkxx}\end{equation}
The equivalence between this propagator and the time-evolution for
the QKR is clear and the parameters correspond as before: $JTB_{Q}\to K$
and $B_{Q}\to\tau$. Therefore, a kicked fermion evolves in position
in the same way as a QKR evolves in momentum i.e. $j\to l\tau$. This
multiple-fermion correspondence is a direct extension of the single-flip
analysis \citep{Boness061,Boness062}. 

Calculating two-particle correlation functions is now straighforward.
For example, we find the matrix element of the Floquet operator in
the two spin-flip basis $|n_{1},n_{2}\rangle=-c_{n_{2}}^{\dagger}c_{n_{1}}^{\dagger}|\phi\rangle$,\begin{small}\begin{eqnarray}
\langle n_{1},n_{2}|U(T)|m_{1},m_{2}\rangle & = & \langle\phi|c_{n_{1}}c_{n_{2}}U(T)c_{m_{1}}^{\dagger}c_{m_{2}}^{\dagger}|\phi\rangle\\
 & = & \left(\langle\phi|c_{m_{1}}c_{m_{2}}c_{n_{2}}^{\dagger}(T)c_{n_{1}}^{\dagger}(T)|\phi\rangle\right)^{*}.\nonumber \end{eqnarray}
\end{small}Substituting \eqref{eq:fevkxx} into this\begin{small}\begin{equation}
\begin{split}\langle n_{1},n_{2}|U(T)|m_{1},m_{2}\rangle=\sum_{i_{1},i_{2}}e^{-i\frac{B_{Q}}{2}\left[\left(n_{1}-j_{0}\right)^{2}+(n_{2}-j_{0})^{2}\right]}\times\\
i^{i_{1}+i_{2}-n_{1}-n_{2}}J_{i_{1}-n_{1}}(-JT)J_{i_{2}-n_{2}}(-JT)\times\\
\langle\phi|c_{m_{1}}c_{m_{2}}c_{i_{2}}^{\dagger}c_{i_{1}}^{\dagger}|\phi\rangle.\end{split}
\end{equation}
\end{small}From Wick's theorem $\langle\phi|c_{m_{1}}c_{m_{2}}c_{i_{2}}^{\dagger}c_{i_{2}}^{\dagger}|\phi\rangle=\delta_{m_{1},i_{1}}\delta_{m_{2},i_{2}}-\delta_{m_{1},i_{2}}\delta_{m_{2},i_{1}}$
and therefore,\begin{small}\begin{equation}
\begin{split}\langle n_{1},n_{2}|U(T)|m_{1},m_{2}\rangle=e^{-i\frac{B_{Q}}{2}\left[\left(n_{1}-j_{0}\right)^{2}+(n_{2}-j_{0})^{2}\right]}\times\\
i^{n_{1}+n_{2}-m_{1}-m_{2}}\left[J_{m_{1}-n_{1}}(-JT)J_{m_{2}-n_{2}}(-JT)\right.\\
\left.-J_{m_{1}-n_{2}}(-JT)J_{m_{2}-n_{1}}(-JT)\right].\end{split}
\end{equation}
\end{small}Substituting $J_{j}(-\beta)=J_{-j}(\beta)$ into this
yields \eqref{eq:XXTMPROP}.

Finally, we note that when $\Delta\neq0$ the Ising term, $H_{ZZ}=-J\Delta\ \sum_{n}\sigma_{n}^{Z}\sigma_{n+1}^{Z}/4$,
introduces a mutual interaction between the fermions: \begin{equation}
H_{\textrm{ZZ}}=J\Delta\sum_{j=1}^{N}c_{j}^{\dagger}c_{j}-\mathbb{I}/4-c_{j}^{\dagger}c_{j+1}^{\dagger}c_{j+1}c_{j}.\label{eq:HMI}\end{equation}
As a result, the corresponding QKR images will be coupled.


\begin{thebibliography}{31}
\bibitem{Zurek03} W.H. Zurek, Rev. Mod. Phys. \textbf{75}, 715 (2003). 

\bibitem{Benenti07} G.Benenti, G. Casati, and G. Strini, {\em Principles
of Quantum Computation and Information}, Vol 2 (World Scientific,
2007). 

\bibitem{rossini:036209} D. Rossini, G. Benenti, and G. Casati, Phys.
Rev. E \textbf{74}, 036209 (2006). 

\bibitem{Zurek94} W.H. Zurek and J.P. Paz, Phys. Rev. Lett. \textbf{72},
2508 (1994). 

\bibitem{Miller1999} P.A. Miller and S. Sarkar, Nonlinearity \textbf{12},
419 (1999). 

\bibitem{Miller1999a} P.A. Miller and S. Sarkar, Phys. Rev. E \textbf{60},
1542 (1999). 

\bibitem{casati1979} G. Casati, B.V. Chirikov, F.M. Izraelev and
J. Ford, Lect. Notes in Phys. \textbf{93} (Springer-Verlag, New York,
1979) p334. 

\bibitem{Moore} F. Moore, J.C. Robinson, C. Bharucha, B. Sundaram,
and M.G. Raizen, Phys. Rev. Lett \textbf{75}, 4598 (1995); M.G. Raizen,
Adv. At. Mol. Opt. Phys. \textbf{41}, 43 (1999). 

\bibitem{Deco} H. Ammann, R. Ray, N. Christensen and I. Shvarchuck,
J. Phys. B \textbf{31}, 2449 (1998). 

\bibitem{kickatom} P. Szriftgiser, J. Ringot, D. Delande, J.-C. Garreau,
Phys. Rev. Lett. \textbf{89}, 224101 (2002); P.H. Jones, M. Stocklin,
G. Hur, and T.S. Monteiro, Phys. Rev. Lett. \textbf{93}, 223002 (2004);
C. Ryu, M.F. Andersen, A. Vaziri, M.B. d'Arcy, J.M. Grossman, K. Helmerson,
and W.D. Phillips, Phys. Rev. Lett. \textbf{96}, 160403 (2006); M.
Sadgrove, M. Horikoshi, T. Sekimura, and K. Nakagawa, Phys. Rev. Lett.
\textbf{99}, 043002 (2007); I.Dana, V. Ramareddy, I. Talukdar and
G. S. Summy, Phys. Rev. Lett. 100, 024103 (2008); J. F. Kanem, S.
Maneshi, M. Partlow M. Spanner and A. M. Steinberg Phys. Rev. Lett.
98, 083004 (2007). 

\bibitem{Boness061} T. Boness, S. Bose and T.S. Monteiro, Phys. Rev.
Lett. \textbf{96}, 187201 (2006).

\bibitem{Boness062} T. Boness, M. Stocklin and T.S. Monteiro, Prog.
Theor. Phys. Suppl. \textbf{166}, 85 (2007).

\bibitem{Fujisaki2003} S. Adachi, M. Toda and K. Ikeda, Phys. Rev.
Lett. \textbf{61}, 659 (1998); H. Fujisaki, A. Tanaka and T. Miyadera,
J. Phys. Soc. Jpn. Suppl. C \textbf{72}, 111 (2003). 

\bibitem{Park2003} H.K. Park and S.W. Kim, Phys. Rev. A \textbf{67},
060102 (2003). 

\bibitem{LAK2001} A. Lakshminarayan, Phys. Rev. E \textbf{64}, 036207
(2001). 

\bibitem{Wood1990} B.P. Wood, A.J. Lichtenberg, and M.A. Lieberman,
Physica D \textbf{71}, 132 (1994). 

\bibitem{Nag} S. Nag, G. Ghosh and A. Lahiri, Physica D \textbf{204},
110 (2005).

\bibitem{Gong2007} J. Gong and P. Brumer, Phys. Rev. A \textbf{75},
032331 (2007). 

\bibitem{Prosen} T.Prosen, Phys. Rev. E \textbf{60}, 1658 (1999);
Phys. Rev. E \textbf{65}, 036208 (2002). 

\bibitem{Fishman1982} S. Fishman, D.R. Grempel and R.E. Prange, Phys.
Rev. Lett \textbf{49}, 509 (1982). 

\bibitem{Grempel1984} D.R. Grempel, R.E. Prange and S. Fishman, Phys.
Rev. A \textbf{29}, 1639 (1984). 

\bibitem{Kudo} K. Kudo and T.S. Monteiro, Phys. Rev. E \textbf{77},
055203(R) (2008). 

\bibitem{Hanson} J.D. Hanson, E. Ott, and T.M. Antonsen, Phys. Rev.
A \textbf{29}, 819 (1984). 

\bibitem{KarbMull98} H. Bethe, Z. Phys. \textbf{71}, 205 (1931);
M. Karbach and G. Müller, Computers in Physics \textbf{11}, 36 (1997).

\bibitem{Takahashi} M. Takahashi, {\em Thermodynamics of One-Dimensional
Solvable Models}, {Cambridge (2005)}.

\bibitem{Amico04} L. Amico, A. Osterloh, F. Plastina, R. Fazio and
G.M. Palma, Phys. Rev. A \textbf{69}, 022304 (2004); L. Amico, R.
Fazio, A. Osterloh and V. Vedral, Rev. Mod. Phys. \textbf{80}, 517-576
(2008). 

\bibitem{Sub04} V. Subrahmanyam, Phys. Rev. A \textbf{69}, 034304
(2004). 

\bibitem{Santos03} L.F. Santos and M.I. Dykman, Phys. Rev. B \textbf{68},
214410 (2003).

\bibitem{Shepelyansky94} D.L. Shepelyansky, Phys. Rev. Lett. \textbf{73},
2607 (1994). 

\bibitem{Jordan} P. Jordan and E. Wigner, Z. Phys. \textbf{47}, 631
(1928); E. Lieb, T. Schultz and D. Mattis, Ann. Phys. \textbf{16},
406 (1961). 
\end{thebibliography}
\end{document}